\documentclass[aps,12pt,paper]{revtex4-2}
\usepackage{subcaption}
\usepackage{graphicx}
\usepackage{dcolumn}
\usepackage{bm}
\usepackage{xcolor, soul}
\usepackage{amsfonts}
\usepackage{booktabs}
\usepackage{siunitx}
\usepackage{ragged2e}
\usepackage[colorlinks,citecolor=red,urlcolor=blue,bookmarks=false,hypertexnames=true]{hyperref} 
\usepackage{float}
\usepackage{float}
\usepackage{xcolor}
\usepackage{array}
\usepackage{setspace}
\usepackage{lipsum}

\newcommand{\HM}[1]{{\color{black}{#1}}}

\newcolumntype{P}[1]{>{\centering\arraybackslash}p{#1}}
\newcolumntype{M}[1]{>{\centering\arraybackslash}m{#1}}
\usepackage{setspace}
\usepackage{amsmath}
\usepackage{parskip}

\begin{document}

\title{ 
Flow of a two-dimensional liquid foam: Impact of surfactant type and boundary conditions}
\noindent
\author{Farshad Nazari}
\altaffiliation{Department of Chemical and Biomedical Engineering, FAMU-FSU College of Engineering, Tallahassee, FL, 32310, USA}
\author{Andrei Potanin}
\altaffiliation{Colgate-Palmolive Company, Piscataway, NJ 08854, USA}
\author{Hadi Mohammadigoushki}
\altaffiliation{Department of Chemical and Biomedical Engineering, FAMU-FSU College of Engineering, Tallahassee, FL, 32310, USA}
\email[Corresponding Author: ]{hadi.moham@eng.famu.fsu.edu}

\date{\today}


\begin{abstract}

In this study, we experimentally investigate the rheological and flow behavior of two-dimensional (2D) monodisperse aqueous foams, sheared between parallel plates using a custom-made rheo-optical apparatus with smooth and roughened walls. The foams were prepared using two commercially available detergents—Foam 1 and Foam 2—while maintaining similar bubble sizes and liquid fractions. The linear viscoelastic results reveal that the Foam 1 consistently exhibits higher elastic and loss moduli than the Foam 2 , regardless of boundary conditions, with roughened walls further enhancing these moduli in both foams. Additionally, the Foam 1  shows a lower viscoelastic relaxation frequency compared to the Foam 2, indicating a less mobile interface for the Foam 1. In the non-linear regime, significant differences were observed. Under smooth boundary conditions, Foam 2 exhibits yield stress behavior, whereas Foam 1 does not, despite having higher viscous stresses. The viscous stress in the Foam 1 scales with the capillary number as $\tau_w \propto$ Ca$^{0.5}$, while for the Foam 2, the scaling depends on the boundary conditions: $\tau_w \propto$ Ca$^{0.85}$ for smooth walls and $\tau_w \propto$ Ca$^{0.65}$ for roughened walls. These variations in rheological behavior are attributed to differences in surfactant chemistry, leading to different interface mobilities, with the Foam 1 having a less mobile interface compared to the Foam 2.  
\end{abstract}

\maketitle
\newpage 
\section{Introduction}

{Liquid foams consist of a dispersion of gas bubbles in a liquid medium. Despite these two constituents being simple and Newtonian, their combination creates a complex fluid that exhibits strong non-Newtonian rheology. The rheology of the liquid foam depends on multiple factors, including the liquid content, bubble size, bubble size distribution among other factors~\cite{dollet2014rheology,cohen2013flow,denkov2020physicochemical,hohler2005rheology}. Liquid foams are widely used in a host of applications in oil-gas industry~\cite{conn2014visualizing,wei2018foam,salonen2012dual}, mining~\cite{somasundaran1972foam}, cosmetic~\cite{arzhavitina2010foams,parsa2019foam} and house-hold products~\cite{zocchi2008foam}. } { The prior published literature on the rheology of liquid foams is expansive and there are numerous excellent reviews on this subject matter, which the readers are referred to \cite{dollet2014rheology,cohen2013flow,denkov2020physicochemical,hohler2005rheology}. Liquid foams possess high interfacial energy, making them unstable. Early studies investigated the stability of liquid foams in the quiescent state. In principle, three mechanisms could affect the foam structure; drainage due to gravity, Ostwald ripening due to the pressure gradient between neighboring bubbles, and coalescence due to the rupture of the thin film between neighboring bubbles~\cite{durian1991scaling,durian1995foam,cohen2001bubble,saint2006physical,weaire1999physics}. \par 

Under flow, stable liquid foams exhibit a range of complex and rich dynamics~\cite{cohen2013flow,denkov2012foam}. The rheology of liquid foam is strongly dependent on the structure of the liquid foam (e.g., bubble size, bubble size distribution, liquid content, etc.)~\cite{hohler2005rheology}. Typically, under small deformations (linear viscoelastic regime), liquid foams exhibit a strong solid-like behavior (that is, elasticity) due to the presence of liquid interfaces between bubbles~\cite{cohen2013flow}. Under strong and non-linear deformation rates, liquid foam has been known to exhibit yield stress properties at low shear rates and viscous dissipation at higher rates~\cite{reinelt1990shearing,ovarlez2010investigation,dollet2014rheology}. These complex rheological features have been reported to be related to local re-arrangement of bubble dynamics ~\cite{dollet2014rheology,langevin2014rheology,buzza1995linear}. In addition, the rheological response of liquid foams has been shown to be strongly impacted by the type and chemistry of the surfactants used to make liquid foams ~\cite{denkov2009role,marze2008aqueous,golemanov2008breakup,denkov2005wall,golemanov2008surfactant}. Typically, the steady-state flow curve of the liquid foam in non-linear viscoelasticity regime can be described by a Herschel-Bulkley fluid model: 
\begin{equation}
   \tau (\dot{\gamma}) = \tau_y + \tau_v = \tau_y + K{\dot{\gamma}}^n,
   \label{HBmodel}
\end{equation}
where $\tau_y$, $K$, $\dot{\gamma}$, and $n$ are yield stress, consistency factor, imposed shear rate and shear thinning index, respectively.  Experiments on three-dimensional foam samples have shown that, depending on the surfactant chemistry, bubble interfaces can be mobile or immobile (rigid), which is reflected in $n \approx 1/2$ for mobile interfaces and $n\approx 0.2-0.3$ for immobile interfaces when friction between bubble layers is dominant~\cite{denkov2009role}. This scaling is also affected by the type of boundary conditions used to measure the flow curve of liquid foam~\cite{denkov2009role}. For example, if there is significant friction between the bubbles and the wall, the scaling changes to $n \approx 2/3$ for mobile interfaces and to $n \approx 1/2$ for immobile interfaces~\cite{bretherton1961motion,schwartz1986motion,denkov2009role,denkov2005wall}. 
\par

Three-dimensional liquid foams are opaque, making it difficult to link bubble dynamics to rheology. To address this challenge, researchers have used two-dimensional (2D) liquid foams (bubble monolayers) as a model system to investigate the connection between microstructure and rheology~\cite{debregeas2001deformation,lauridsen2004velocity,katgert2008rate,katgert2010couette,dennin2004statistics,wang2006impact,mohammadigoushki2012anomalous,mohammadigoushki2013size}. Experiments on the flow of 2D liquid foams have revealed a range of intriguing characteristics, including non-local rheology~\cite{katgert2008rate}, size-based segregation~\cite{mohammadigoushki2013size,mader2012quantitative,mohammadigoushki2015temporal}, coalescence~\cite{mohammadigoushki2012anomalous}, breakup~\cite{golemanov2008breakup}, among others. Theoretical studies have suggested a shear shear-thinning index of $n \approx 1/2$ for 2D polydisperse foams~\cite{janiaud2006two,langlois2008rheological}. Experiments on 2D foam flow have indicated a shear thinning index of $n = 0.36-0.4$ in a Taylor-Couette cell (flow between two concentric cylinders) without the top wall and $n \approx 2/3$ in a parallel plate geometry for a 2D foam made of a commercial detergent, which is believed to lead to a mobile interface~\cite{katgert2008rate,katgert2010couette}. The latter experiments studied the flow curves of two-dimensional foams within a very low range of shear rates ($\dot{\gamma}< 0.1$ [1/s]). Consequently, the behavior of 2D liquid foams under higher shear rates remains unexplored.\par 

To the best of our knowledge, the impact of surfactant chemistry on non-linear rheology, specifically regarding mobile or immobile interfaces, has not been studied in the flow of 2D liquid foams.  Previous studies on the rheology of 2D foam have mainly utilized only one detergent~\cite{katgert2008rate,katgert2010couette}. Furthermore, while the impact of surfactant chemistry on the non-linear rheological properties of 3D liquid foams has been extensively studied, its effect on the linear viscoelastic properties of foams has received little attention~\cite{krishan2010fast,costa2013coupling}. Krishan et al. showed that for a 3D liquid foam with immobile interfaces, the linear viscoelastic rheology is different from the 3D liquid foam with mobile interface~\cite{krishan2010fast}. In particular, Krishan et al. showed that the viscoelastic relaxation frequency ($f_c$; see details below) is typically higher for mobile interfaces than for immobile interfaces~\cite{krishan2010fast}. In addition, while the elastic modulii of the samples with immobile and mobile interfaces were similar, the loss modulus obtained for the 3D liquid foam with immobile interfaces is larger than that obtained for liquid foams with mobile interfaces~\cite{krishan2010fast}. It remains unclear how or whether the type or properties of the surfactant influences the linear viscoelastic rheological properties of 2D liquid foams.\par

Finally, most experiments on liquid foams have focused on characterizing the rheology of stable foams. There is limited research on flow-induced microstructural changes in liquid foams, such as breakage~\cite{golemanov2008breakup} and size-based segregation~\cite{mohammadigoushki2013size,mohammadigoushki2012anomalous,mohammadigoushki2014bubble}. Consequently, little is known about how flow affects the microstructure of 2D foams and its potential feedback effects. \HM{Our study aims to address key knowledge gaps in the literature regarding the flow of 2D liquid foams. Specifically, we will examine how surfactant chemistry, which dictates interfacial mobility (mobile vs. immobile interfaces), and different boundary conditions (smooth vs. roughened) influence both linear and nonlinear rheology. Additionally, we will investigate the potential structural effects these factors impose on 2D liquid foams. } To accomplish this, we will employ two industrially relevant surfactant systems and utilize a custom-designed rheo-optical cell. This setup features flow between two parallel plates and an in situ-generated monodisperse foam, enabling direct visualization of foam structure through a tailored optical system.}

\section{Materials and Methods}
\HM{\subsection{Setup Description }}
The two-dimensional (2D) foam is generated using a dilute (0.1 wt$\%$) solution of commercially available dish soaps, and are called Foam 1 and Foam 2. These liquid foams are made by using two commercially available dishwashing detergents. \HM{The surface tension of the detergent solutions was determined using the contact angle method, which measures the angle formed between a sessile droplet and a glass substrate. Surface tension, $\gamma$, is calculated as $\gamma = {F}/{L \cdot \cos \theta}$, where $F$ is the force acting on the plate, $\theta$ is the contact angle, and $L$ is the wetting length.  Droplets of detergent solution with a fixed volume were dispensed onto the glass substrate using a syringe pump and needle. The contact angle and wetting length were measured for various detergent concentrations using an optical setup equipped with a high-resolution camera and analyzed via the "Drop Analysis-LB-ADSA" plugin in ImageJ software.  The force was determined by measuring droplet mass with a precision balance accurate to $10^{-5}$ grams.}\par 

\begin{figure}[ht]
\includegraphics[width=0.99\linewidth]{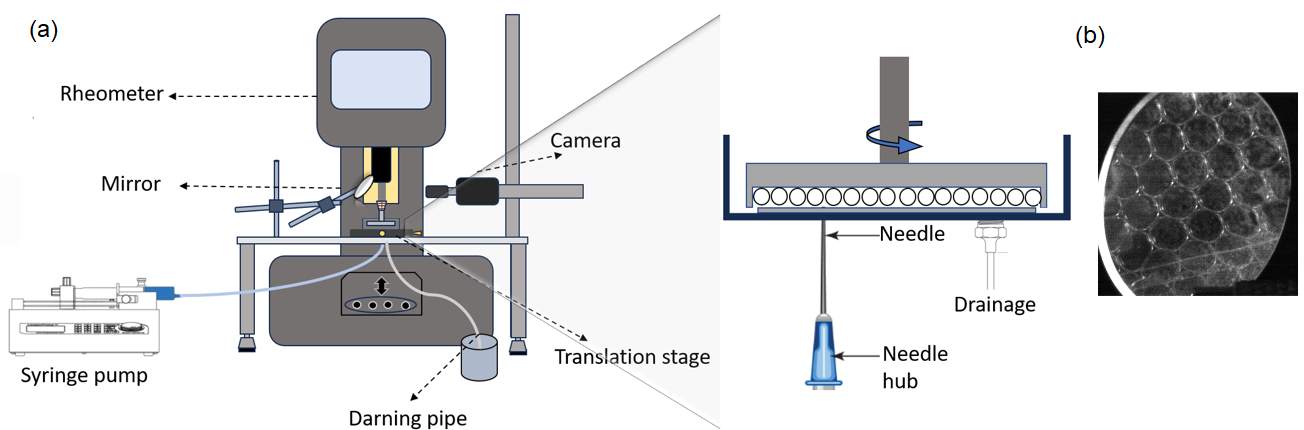}
\caption{\small A schematic of the experimental setup used to generate and study the flow and structure of the 2D liquid foams. The top plate is either a sand-blasted acrylic sheet or roughened using the sand paper. (b) a snapshot of the liquid foam generated between the two walls.}
\label{setup} 
\end{figure}

As shown in Fig.~\ref{setup}, the 2D liquid foam is created by injecting air through the bottom of the lower plate using a \HM{stainless steel} needle \HM{(with a 0.01 inch inner diameter)} and a syringe pump \HM{ (from Sigma-Aldrich)}. The size of the bubbles is controlled by adjusting the \HM{ambient} air flow rate through the needle and the syringe pump. In addition, the liquid content is regulated using a drainage valve at the bottom of the cup \HM{(with a 0.2 inches in diameter)}. The 2D liquid foam is sandwiched between two parallel plates of 63.5 mm in diameter, and sheared using a TA-HR10 commercial rheometer \HM{(TA Instruments)}. The gap between the two plates is maintained at 1mm. We use three different types of boundary conditions (BC). \HM{1- SNC (which stands for sandblasted with no cap; see Fig. S1(c) in the suppelementary materials): The top and bottom plates are sandblasted using a compressed air-driven sandblasting machine (from McMaster-Carr) equipped with a nozzle of (5/16 inches in dimeter). The abrasive material used was silica sand with a grain size of  typically between 10-20 $\mu m$. To achieve uniform roughness, the plates were placed inside a sandblasting chamber, and the abrasive particles were propelled at the surface under a pressure of (80 psi). The nozzle was kept at a consistent distance of (30 cm) from the plate surface to ensure even roughening across the entire area. The plates were rotated and adjusted to cover all sections uniformly. 2- SC (which stands for sandblasted with cap; see Fig. S1(d) in the suppelementary materials): The plates are sandblasted but enclosed with a cover cap to minimize evaporation and maintain consistent humidity. 3- RC (which stands for roughened with cap; see Fig. S1(e) in the suppelementary materials): The plates are glued with a grid 50 sandpaper, introducing surface roughness to further enhance plate-foam adhesion and reduce wall-slip.} \HM{A window was left open as an observation area for monitoring the foam structures}.

\HM{\subsection{Experimental Protocol}}

\HM{The rheo-optical cell was first calibrated by measuring the viscosity of a known Newtonian fluid (silicone oil) using SNC and RC boundary conditions and compared with a standard off the shelf concentric cylinders geometry as shown in Fig.~S2 of the supplementary materials. }

\HM{\subsubsection{Liquid Foam Preparation}}
\HM{Before generating the liquid foam, the measuring cell was filled with around 25 mL of the detergent solution with a concentration of 0.1 wt$\%$ in de-ionized water. To generate the 2D foam, the ambient air flows through the needle, (positioned vertically below the lower plate) with a constant flow rate (1-200 [mL/h]). Upon air injection, a single bubble forms at the needle tip due to surface tension at the air-liquid interface. The size of the bubble is controlled by the balance between the air pressure from the syringe pump and the surface tension of the liquid film formed by the dish soap solution (see sample monodisperse foams with different bubble sizes generated by this method in the Fig.~S3 of the supplementary materials). Once the bubble reaches a critical size, where the buoyant force exceeds the surface tension holding it at the needle, the bubble detaches from the needle tip and rises through the liquid towards the gap between the two parallel plates. After detaching from the needle, the bubble becomes trapped against the upper plate. Once the upper plate was fully covered with the desired mono-dispersed bubbles, the knifed edge of the upper plate helps capture and holds the bubble in place, ensuring it remained positioned beneath the upper plate and at this point, the drainage valve at the bottom of the setup is opened to remove excess liquid. During the same time, the lower plate is gradually lowered to create a 1 mm gap between the two parallel plates}. Once the 2D foam is generated between the plates, the bubble size and liquid content (or foam quality) are monitored \HM{using a hight resolution camera (model Basler ace acA800-510um) coupled with a 35mm Canon lens focused on a mirror (50mm in diameter) that was orientated with angle $45^{\circ}$ on the top of the upper plate} over a one hour time frame \HM{and analyzed using ImageJ software} to ensure the stability of the liquid foam in the absence of flow \HM{(see Fig. S4 in the suppelementary materials for sample foam structures over the course of time)}. 

 \HM{\subsubsection{Dynamic Measurements:}}
 Subsequently, two types of rheological measurements are performed. The first experiments involve small-amplitude oscillatory shear experiments in the linear viscoelastic regime. Secondly, steady shear experiments are performed to measure the flow curve for the 2D foams. Additionally, the surface dilatational modulus of the soap solutions and the resulting surface tensions at the air-liquid interface were measured using an oscillating pendant drop method. \HM{Each experiment lasted approximately 40 minutes and was repeated at least three times with fresh samples, resulting in standard deviations below 5\%.} \par  


\section{Results}
\subsection{Foams stability}
\begin{figure}[hthp]

\includegraphics[width=0.48\linewidth]{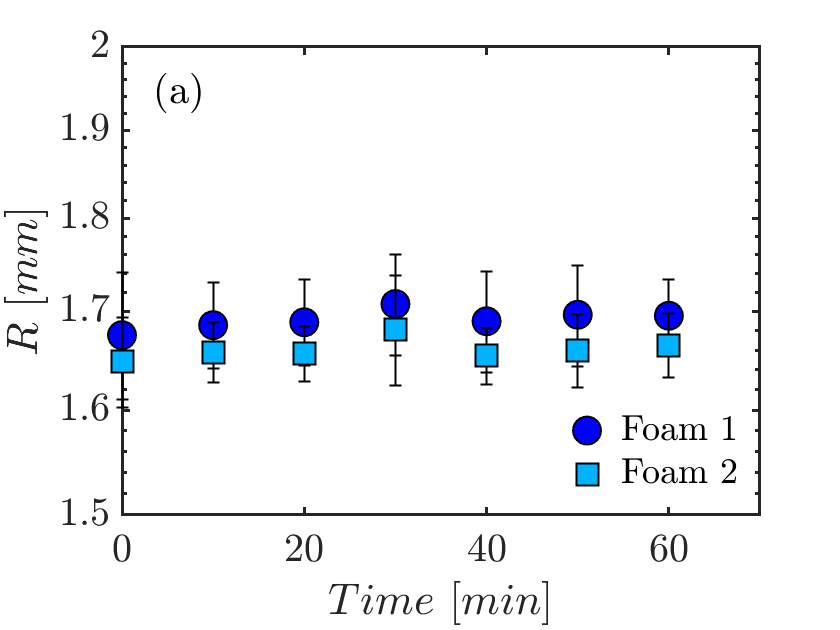}
\includegraphics[width=0.48\linewidth]{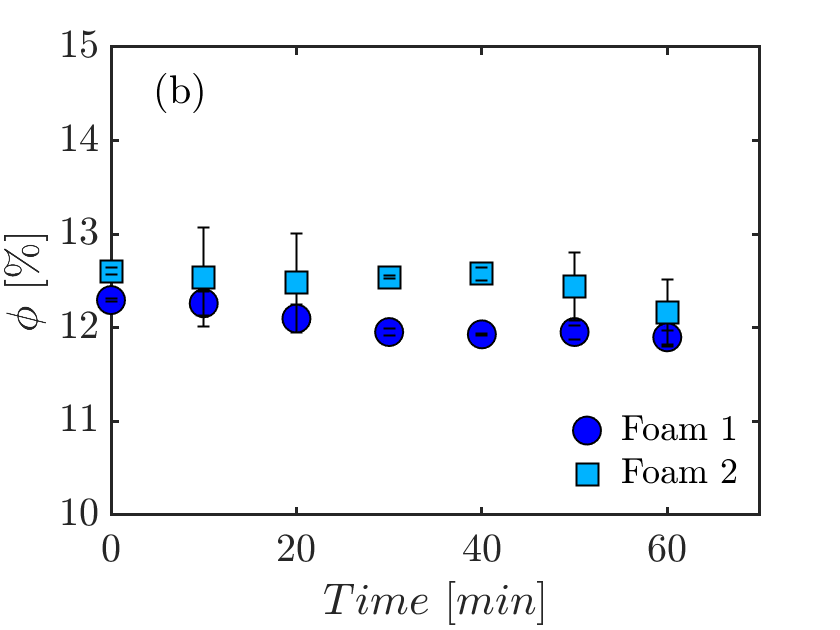}

\caption{\small \HM{The bubble radius (a) and liquid content $\phi$ (b) as a function of time for foams made of two different detergents.} }
\label{fig:stability}
\end{figure}

Before reporting the rheological properties of the two-dimensional foam, we assessed the stability of the liquid foams in a quiescent state. In these experiments, we monitored the foam structure and quantified bubble size and foam quality (liquid content) over time under various boundary conditions. Figure~\ref{fig:stability}(a,b) present the bubble size and liquid fractions ($\phi$) for the two liquid foams used for this study. In particular, the bubble size and the liquid content are similar for these two liquid foams and remain unchanged within the first hour, indicating that the liquid foams are stable. We monitored the foam structure over a couple of hours and observed no significant changes in its morphology. In addition, our results show that the type of boundary condition does not affect the foam structure and stability under no flow conditions.\par  

\subsection{Linear Viscoelastic Rheology}

\begin{figure}[ht]
\centering
\includegraphics[width=0.49\linewidth]{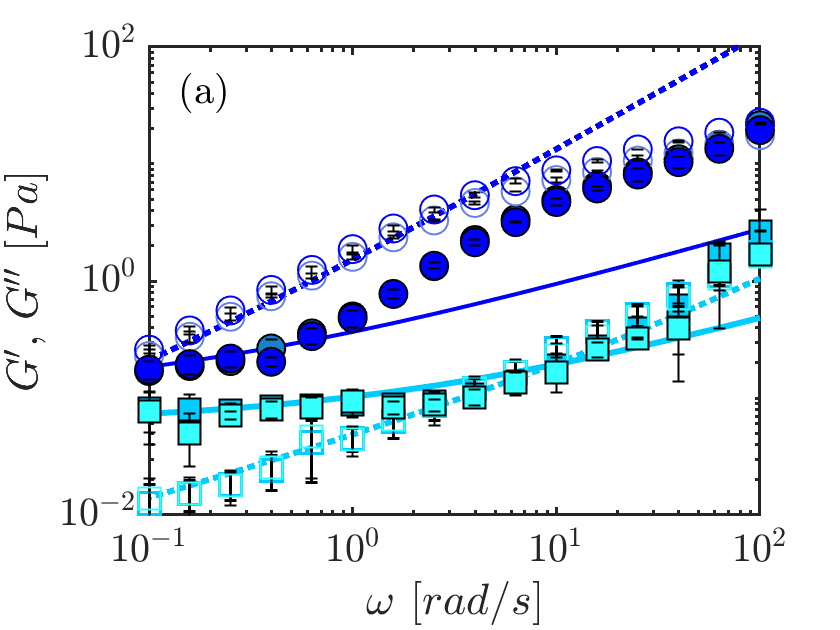}
\includegraphics[width=0.49\linewidth]{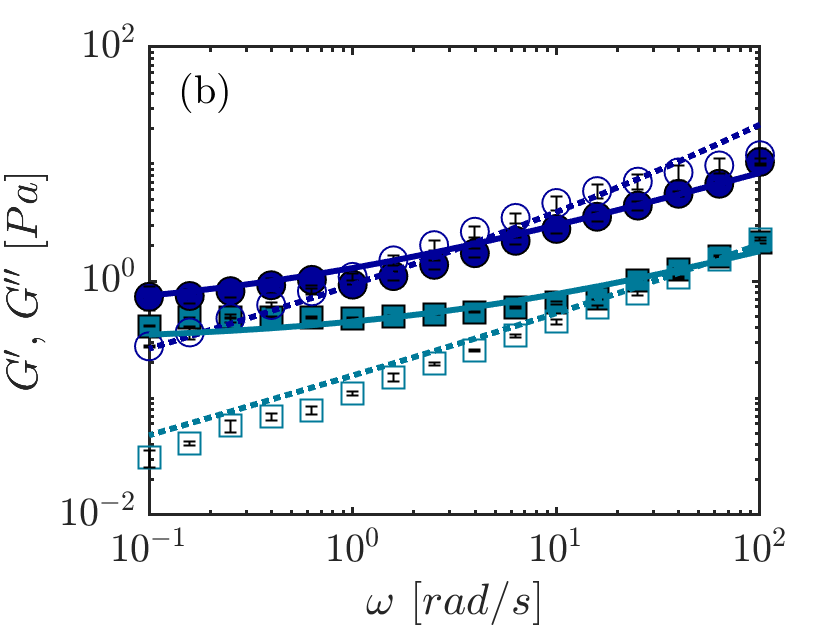}
\caption{\small Storage and loss moduli as a function of angular frequency for (a) no cap-smooth, cap-smooth and (b) cap-roughened walls. The open and closed symbols denote the elastic and loss moduli, respectively. In part (a), the SNC and SC data overlap. Circles denote Foam 1 sample and Squares are for Foam 2. The lines are the best fits based on equation \ref{complex}.}
\label{SAOS}
\end{figure}

Fig.~\ref{SAOS}(a) shows the elastic and loss moduli measured as a function of the angular frequency for the two liquid foam samples when the walls are sandblasted. These measurements are performed with a strain of 1$\%$ to ensure that experiments are performed under linear viscoelastic regime. Below and at this strain, the elastic and loss moduli do not depend on the strain amplitude. The linear viscoelastic response of liquid foams exhibits several qualitative differences in various aspects. First, the 2D liquid foam sample based on Foam 1 shows stronger elastic and loss moduli compared to the foam made by the Foam 2 detergent (see Fig.~\ref{SAOS}(a)). Secondly, the elastic modulus for Foam 2 sample reaches a plateau at low frequencies and is stronger than the loss modulus at low frequencies. Conversely, for the 2D liquid foam based on Foam 1, the elastic modulus is lower than the loss modulus over the entire range of frequencies probed. By adding the cap, the frequency sweep response of two liquid foams remains similar. 

Fig.~\ref{SAOS}(b) shows a comparison between the linear viscoelastic moduli of the 2D liquid foams with roughened boundary conditions. In this case, the moduli for both liquid foam systems have increased significantly compared to the sandblasted (or smooth) boundary condition (c.f. Fig.~\ref{SAOS}(a) and Fig.~\ref{SAOS}(b)). A similar change is observed for the liquid  Foam 2. Although the concentrations of surfactant, liquid content, and bubble size are the same for these two liquid foams, the elastic and loss moduli of Foam 1 are consistently higher than those of the Foam 2. This behavior is observed for all different boundary conditions, which suggests that the type of surfactant may have led to these differences. To better understand the nature of this difference, we performed non-linear shear rheology experiments.




\subsection{Non-Linear Shear Rheology}
\begin{figure}[hthp]
\centering
\includegraphics[width=0.49\linewidth]{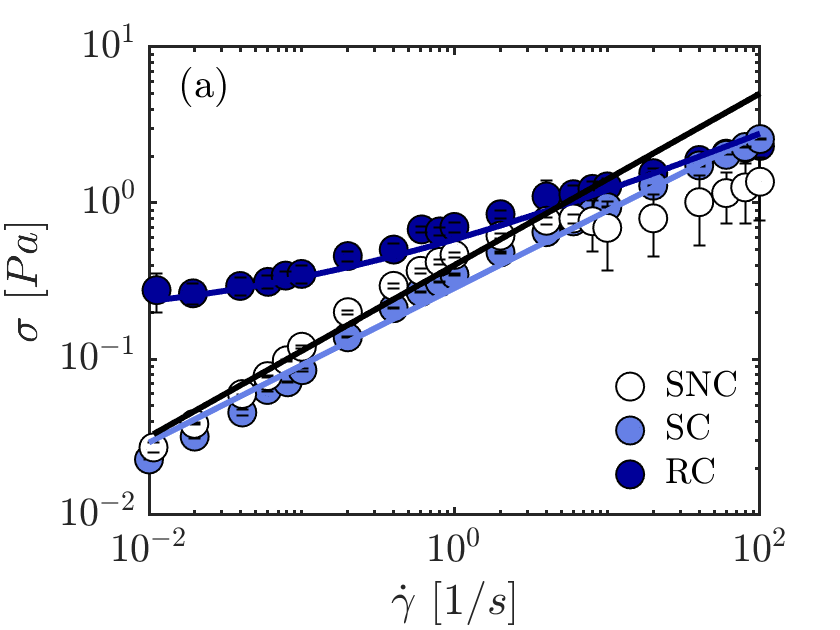}
\includegraphics[width=0.49\linewidth]{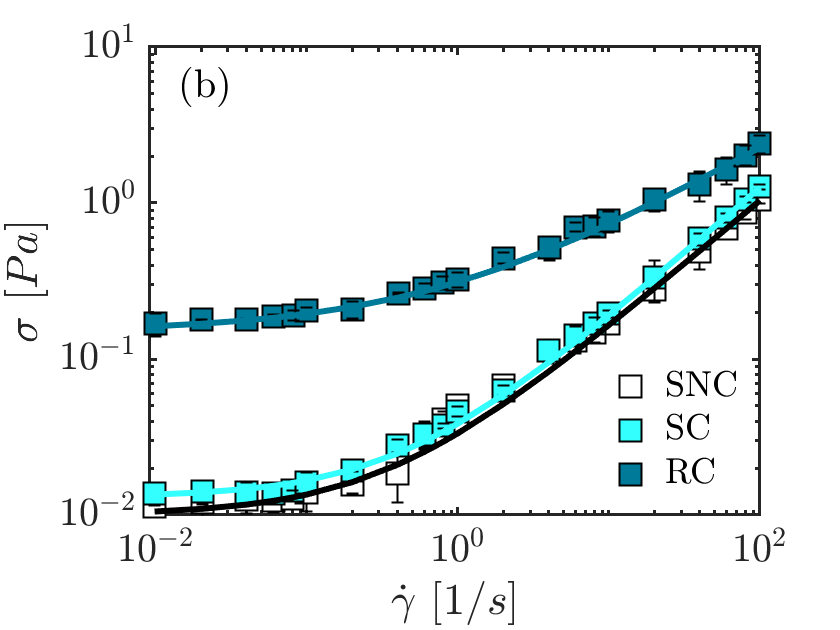}
\caption{\small Steady state flow curve for the two-dimensional liquid foam made of (a) Foam 1 and (b) Foam 2 surfactants. The lines are the best fits based on Power-Law (a) and Herschel-Bulkley (b) models. } 
\label{FC}
\end{figure}

Figure~\ref{FC} shows the measured steady-state shear stress as a function of the applied shear rate for 2D liquid foams made from different surfactant chemistry under different boundary conditions. First, we discuss the effects of boundary conditions on the non-linear rheology of foam samples. Interestingly, the 2D foam made with Foam 1 detergent does not exhibit yield stress behavior in the presence of smooth boundaries (with or without a cap). However, when roughened boundary conditions are introduced, the rheology of the 2D foam changes drastically, displaying a yield stress behavior. In contrast, the 2D liquid Foam 2 exhibits yield stress behavior even with smooth walls (see Fig.~\ref{FC}(b)). When roughened walls (with sandpaper) are used, the measured shear stress for the liquid foams increase significantly. \HM{Interestingly, for nearly all boundary conditions, the shear stress measured for Foam 1 is higher than the shear stress of Foam 2. This is consistent with the results of Fig.~(\ref{SAOS}) that show linear viscoelastic moduli are consistently higher for Foam 1 than Foam 2. }The flow curves are fitted to a Herschel-Bulkeley fluid model and the rheological properties of the liquid foams are summarized in Table~(\ref{HB}). 

In nearly all non-linear rheological measurements, the shear stress increases monotonically with increasing shear rate. However, one exception occurs in experiments involving liquid Foam 1, where smooth boundary conditions were applied without a cap (SNC). At relatively high shear rates, $\dot{\gamma} \approx$ 10 [1/s], the shear stress decreases as the shear rate increases, before rising again for $\dot{\gamma} \geq$ 20 [1/s]. 
\begin{table}[ht]
\caption{\small{Rheological properties of the 2D liquid foams under various boundary conditions. }
}
\label{T1}
\setlength{\tabcolsep}{5pt} 
\renewcommand{\arraystretch}{1}
\begin{ruledtabular}
\begin{tabular}{cccccc}
   Foam & BCs & $\tau_y$ [Pa] & $K$ [Pa.s$^{n}$] &    $n$ &\HM{$R^2$}\\
\hline\hline
  & SNC & - & 0.33\HM{$\pm$0.05} & 0.52\HM{$\pm$0.03}&\HM{0.95}\\ 
 Foam 1 & SC & - & 0.285\HM{$\pm$0.01} & 0.5\HM{$\pm$0.01} &\HM{0.99}\\
  & RC & 0.2\HM{$\pm$0.018} & 0.4\HM{$\pm$0.007} & 0.48\HM{$\pm$0.04}&\HM{0.99}\\
   \hline
  & SNC& 0.014 \HM{$\pm$0.001} & 0.026\HM{$\pm$0.002}& 0.84\HM{$\pm$0.03}\ &\HM{0.99}\\ 
Foam 2 & SC &0.011\HM{$\pm$0.001}& 0.023\HM{$\pm$0.001} & 0.83 \HM{$\pm$0.02}&\HM{0.99}\\
& RC& 0.15 \HM{$\pm$0.01}  & 0.162\HM{$\pm$0.01}  & 0.6\HM{$\pm$0.05} &\HM{0.99}
\end{tabular} 
\end{ruledtabular}

\label{HB}
\end{table}

\section{Discussion}

Our experiments reveal notable differences between the 2D liquid foams produced using different commercial surfactant systems for smooth and roughened boundary conditions. The first major difference observed is that, despite having a similar bubble size and foam quality, the elastic and loss moduli are consistently higher in the 2D liquid Foam 1 compared to those made with Foam 2, regardless of the type of boundary conditions. Given that the bubble size, liquid fraction and even the viscosity of the soap solutions are similar in these two systems, these results suggest that the chemistry of the surfactants may have led to these differences. Surfactants are known to change the behavior of the interface between air-liquid in the thin film formed between bubbles. Krishan et al. reported the elastic and loss moduli of 3D liquid foams with mobile and immobile interfaces\cite{krishan2010fast}. On the basis of their results the loss moduli for the liquid foam with an immobile interface are higher than for the liquid foam with mobile interface. This study suggests that the Foam 1 may generate an interface that is less mobile (or more rigid) than the interface generated by the Foam 2. To evaluate this hypothesis, we took two approaches; one through the idea of viscoelastic relaxation and alternatively, by measuring the interfacial dilatational modulus of these surfactant systems. \par 

Previous studies have suggested that the linear viscoelastic properties of a liquid foam could be presented as~\cite{liu1996anomalous}:
\begin{equation}
    G^{*} = G (1+\sqrt{i\omega/f_c}) + 2\pi i \eta_{\infty}\omega.
    \label{complex}
\end{equation}
Here $G^*$, $G$, $\omega$, $f_c$ and $\eta_{\infty}$ are the complex modulus, plateau modulus, angular frequency, viscoelastic relaxation frequency, and viscosity of the solution. We have fitted Eq.~(\ref{complex}) to small amplitude oscillatory shear experimental data reported in Fig.~\ref{SAOS} and obtained the parameters $G$, $f_c$ and $\eta_{\infty}$ (Fits are shown as continuous and dashed curves in Fig.~\ref{SAOS}). Table~(\ref{Fc}) shows these parameters for 2D liquid foams based on Foam 1 and Foam 2. Although we fitted Eq.~(\ref{complex}) to all experiments, it is clear that for experiments that involve Foam 1 and smooth boundary conditions, \HM{the general trend in linear viscoelastic data with respect to angular frequency does not follow predictions of Eq.~(\ref{complex}), and therefore, the fit quality is not good} (corresponding to a low coefficient of determination $R^2$). An important observation from this analysis is that the viscoelastic relaxation frequency for the liquid foam based on Foam 1 is smaller than that calculated value for Foam 2. Presumably, the lack of surfactant mobility creates a sustained surface tension gradient that makes the interface to strongly resist against deformation and yields a higher energy dissipation and a slower relaxation process. Therefore, the above analysis suggests that the interface of Foam 1 detergent may be less mobile than the interface of the Foam 2.
\vspace{-0.1cm}
\begin{table}[hthp]
\caption{\label{T1} \small The viscoelastic rheological properties of the liquid foam as obtained by fitting Eq.~\ref{complex} to the small amplitude oscillatory shear data.} 
\begin{ruledtabular}
\setlength{\tabcolsep}{5pt} 
\renewcommand{\arraystretch}{1} 
\begin{tabular}{cccccc}

   Foam Detergent & BCs& G [Pa] & $f_c$ [1/s] &$\eta_\infty$ [Pa.s] &   $R^2$\\
\hline\hline
 2& RC & 0.3\HM{$\pm$0.02}  & 2\HM{$\pm$0.01} & 0.001\HM{$\pm$0.0001}&0.99 \\ 
2 & SC & 0.06\HM{$\pm$0.002} \ & 1\HM{$\pm$0.1} & 0.001\HM{$\pm$0.0001}&0.95 \\ 
\hline 
 1 & RC & 0.5\HM{$\pm$0.02} & 0.2\HM{$\pm$0.01}& 0.008\HM{$\pm$0.0005}&0.98 \\

 1 & SC & 0.1\HM{$\pm$0.005}  & 0.07\HM{$\pm$0.002}   & 0.2\HM{$\pm$0.02} &0.61 \\
\end{tabular} 
\end{ruledtabular}

\label{Fc}
\end{table}

\begin{figure}[hthp]
\includegraphics[width=0.49\linewidth]{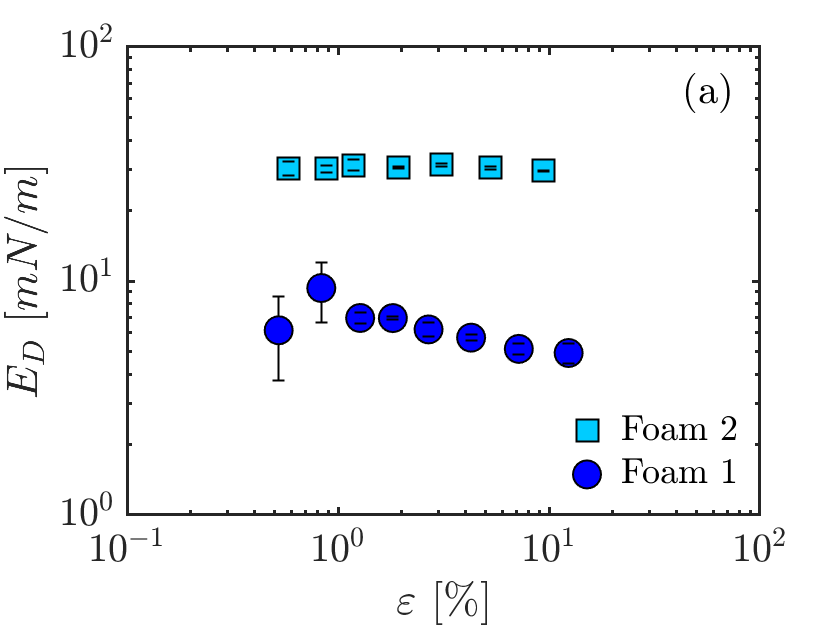}
\includegraphics[width=0.49\linewidth]{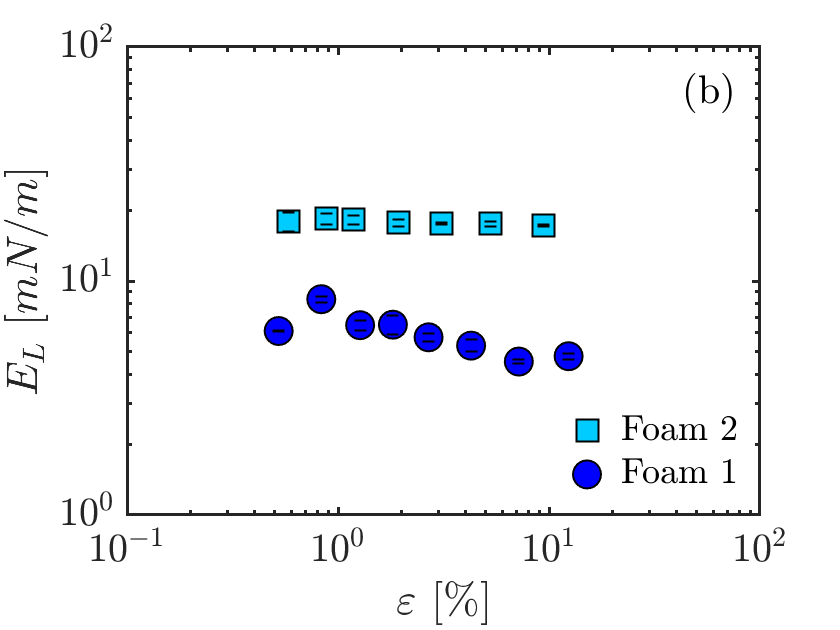}
\caption{\small Surface dilatational modulus $E_{D}$ (a) and the loss surface modulus $E_{LD}$ (b) as a function of strain oscillation for soap solutions made of Foam 1 and Foam 2 detergents. These measurements are performed at a frequency of 1 Hz. }
\label{interRheo}
\end{figure}

To evaluate the liquid-air interface type for the liquid foams used in this study, we performed interfacial rheological measurements using an oscillatory pendant drop tensiometer. In these measurements, the droplet of the solution is perturbed sinusoidally and changes in the surface tension are measured. The changes in surface tension ($\sigma$) can be related to elastic ($E_{SD}$) and loss surface moduli ($E_{LD}$) as below:
\begin{equation}
    \sigma (t) = E_{SD}\varepsilon \sin(\omega t) +  E_{LD}\varepsilon \cos(\omega t).
    \label{Droplet}
\end{equation}
Here $\varepsilon$ denote the imposed strain.  Fig.~\ref{interRheo}(a,b) shows the total surface modulus ($E_{D} = \sqrt{{E_{SD}^2 + {E_{LD}^2}}}$) and the loss surface modulus measured for the two soap solutions as a function of imposed strain oscillations. These measurements indicate that both interfacial moduli are larger for the soap solution based on Foam 2 detergent than for the Foam 1 sample. The latter result is rather surprising because it suggests that the Foam 1 should generate a more mobile interface compared to the Foam 2 surfactant interface, which is at odds with the small-amplitude oscillatory shear data and the above interpretation of the interfaces using the viscoelastic relaxation frequency ($f_c$).  

\begin{figure}[H]
\centering
\includegraphics[width=0.581\linewidth]{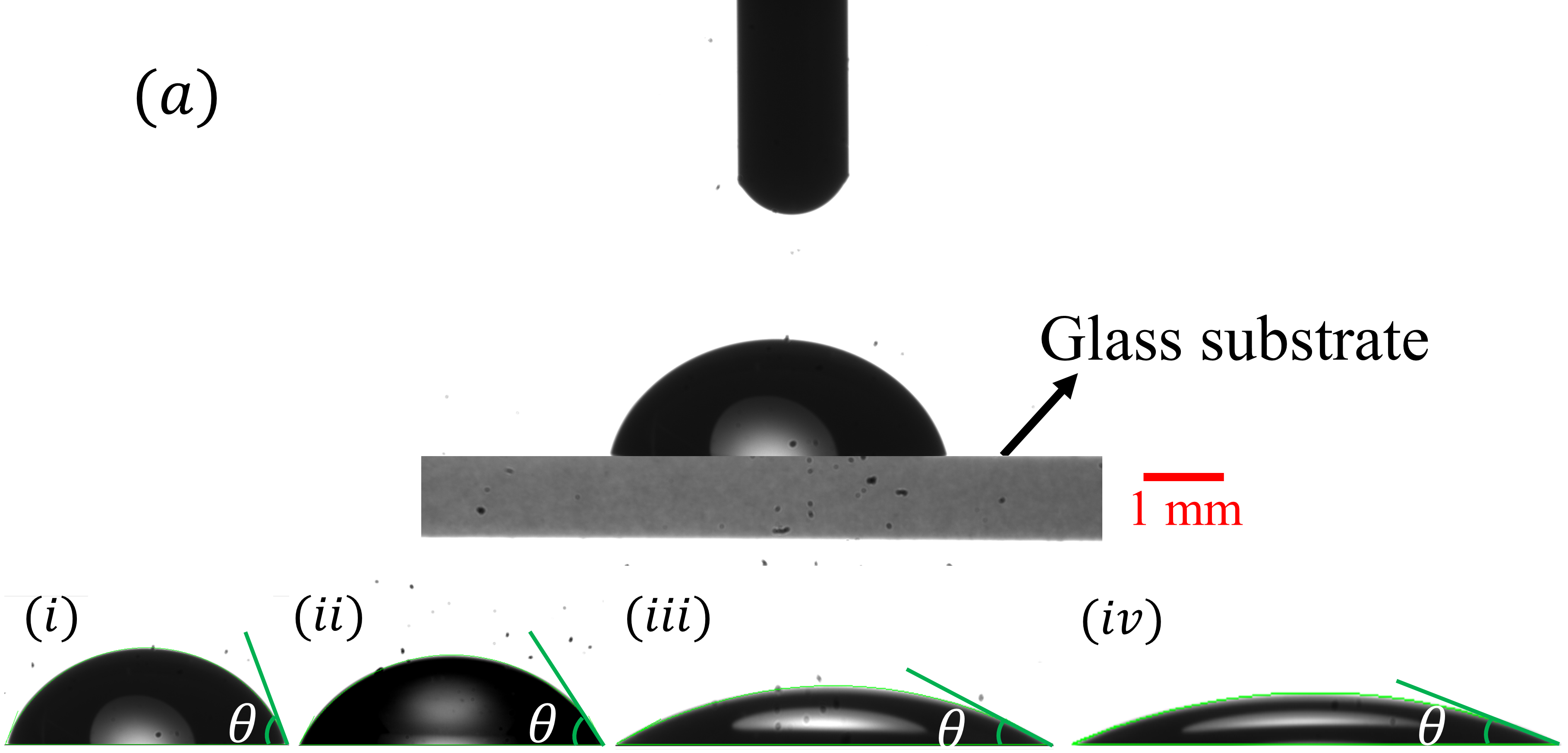}
\includegraphics[width=0.41\linewidth]{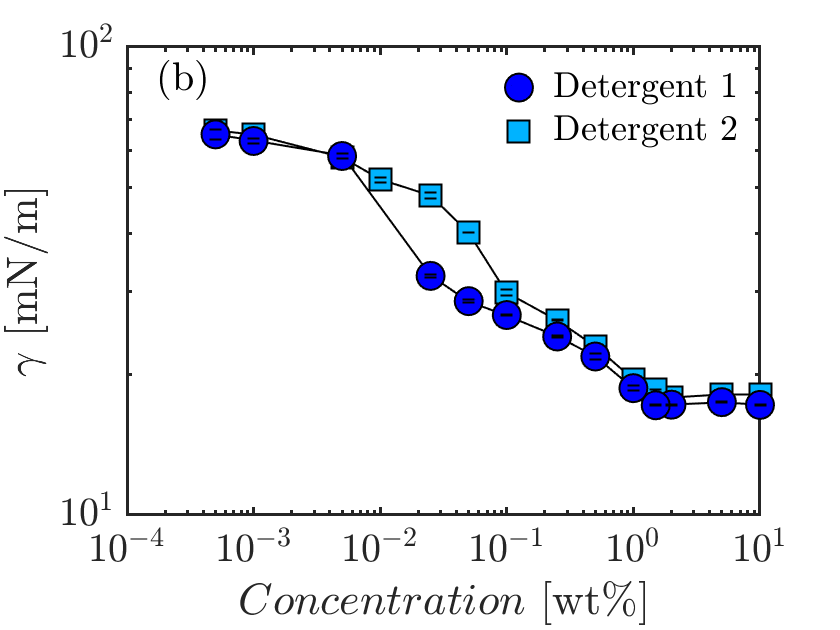}

\caption{\small \HM{ (a) snapshots from contact angle measurements. Sub-figures $(i-iv)$ display sessile droplets on the glass substrate droplets and their corresponding contact angles on the glass substrate for solutions with concentrations of 5$\times10^{-4}$, 5$\times10^{-3}$, 0.5, and 1 [wt\%], respectively. (b) surface tension $(\gamma)$ as a function of detergent concentration for both detergent 1 and detergent 2, across various concentrations. } }
\label{surface tension} 
\end{figure}

\HM{ To further investigate the discrepancies between interfacial dilatational rheology and the linear viscoelastic shear results presented in Fig.~(\ref{SAOS}), we measured the equilibrium surface tension of both detergents as a function of concentration using contact angle measurements. The data of Fig.~\ref{surface tension}(a) shows a series of snapshots taken from the contact angle measurements at various surfactant concentrations. As surfactant concentration increases, the sessile droplet spreads more on the substrate suggesting a reduction in surface tension. Fig.~\ref{surface tension} (b) shows the variation of surface tension as a function of surfactant concentration for the two detergents used in this study. The detergent concentration used to generate 2D liquid foams (0.1 wt\%) falls below the critical micelle concentration (CMC) for both systems (CMC $\approx$ 1.5 wt$\%$ for both systems). With surface tensions of 30 [mN/m] and 26 [mN/m] at 0.1 wt\%, compared to approximately 17 [mN/m] beyond the CMC, these results suggest that the droplet interface in oscillating pendant drop experiments may not be fully saturated with surfactants. When a surfactant-laden interface undergoes dilation or contraction below the CMC, surfactant exchange between the bulk phase and the interface may be too slow to maintain equilibrium (or fully covered interface by surfactants), leading to interfacial concentration gradients and the generation of Marangoni stresses~\cite{langevin2014marangoni,langevin2014rheology,marquez2025measurement}. Marangoni stresses drive interfacial flows from regions of lower to higher surface tension, which may redistribute surfactants and affect interfacial dynamics~\cite{langevin2014marangoni,langevin2014rheology}. In linear viscoelastic (SAOS) measurements, where simple shear is applied without compression or dilation, Marangoni stresses are negligible. However, they play a significant role in dilatational rheology, complicating direct comparisons between shear and dilatational results. We hypothesize that the observed discrepancies between dilational and shear rheology measurements are due to an additional level of complexity associated with Marangoni stress effects in interfacial dilational measurements. Consequently, interfacial rheological data should be interpreted with caution, considering the potential influence of these stresses. Prior studies on liquid foam rheology and surfactant dilatational behavior have largely focused on concentrations above the CMC, where Marangoni effects are minimal~\cite{golemanov2008surfactant,denkov2009role}. To bridge the gap between shear and dilatational rheology, future research should systematically investigate interfacial and foam dynamics across a broad concentration range, encompassing both below and beyond CMC conditions.}

\begin{figure}[hthp]
\centering
\includegraphics[width=0.7\linewidth]{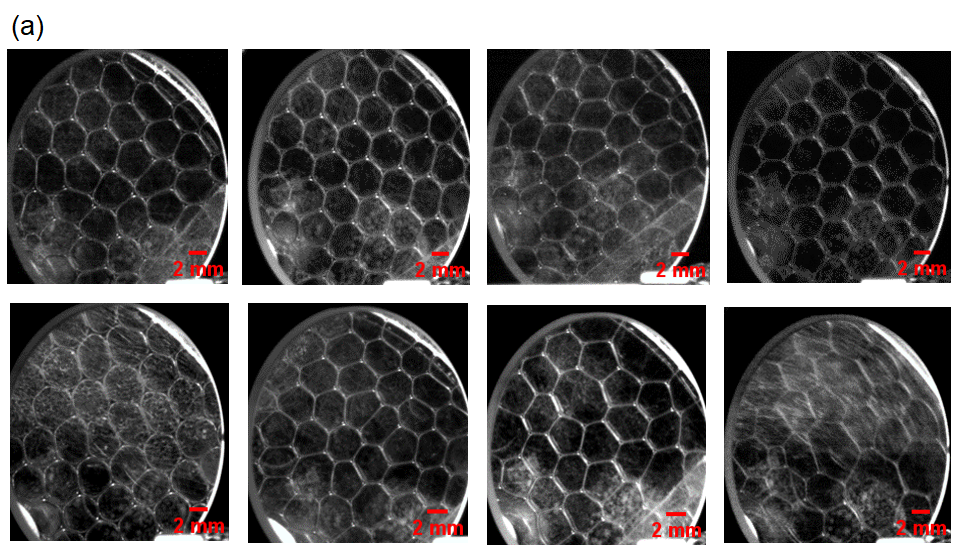}
\includegraphics[width=0.45\linewidth]{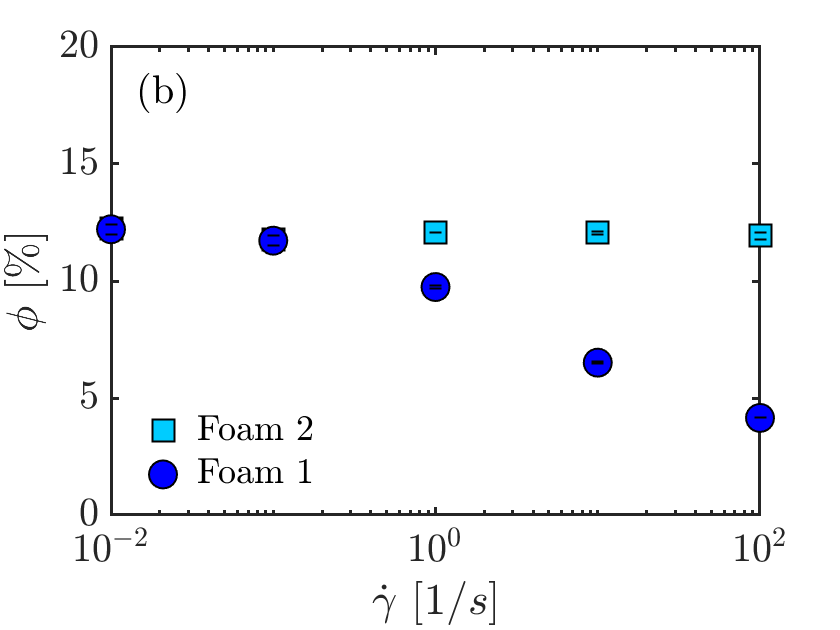}
\includegraphics[width=0.45\linewidth]{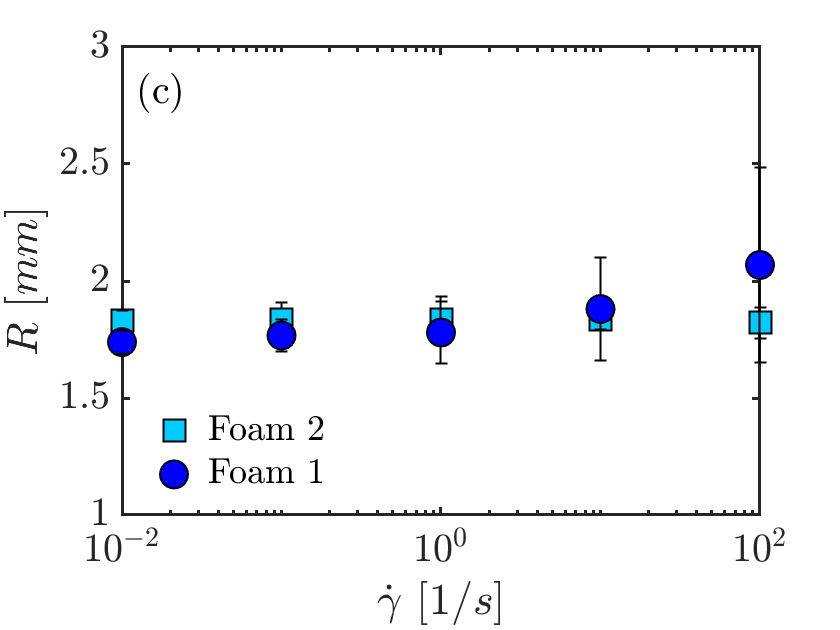}
\caption{\small (a) A series of snapshots taken from the 2D foam samples under different conditions for Foam 2 (top row) and Foam 1 (bottom row). From left to right shear rates are 0.1, 1, 10, and 100 [1/s]. (b) Liquid fraction and (c) averaged bubble radius as function of shear rate for both Foam 1 and Foam 2 samples. }
\label{snapshots}
\end{figure}

Moving beyond the linear viscoelastic results, we now focus on the non-linear viscoelastic response of the 2D liquid foams, beginning with the non-monotonic trend observed in the shear stress data for the Foam 1 with smooth boundary conditions (see Fig.~\ref{FC}(a)). To gain deeper insights into this behavior and how it differs from other conditions, we monitored the evolution of the foam structure during flow imposition across all liquid foam samples. Fig.~\ref{snapshots}(a) presents a series of snapshots that compare Foam 2 (top row) and Foam 1 (bottom row) for smooth boundary conditions. Despite similar initial conditions, Foam 1 undergoes flow-induced coalescence at high shear rates ($\dot{\gamma} \geq 10$), leading to the formation of larger bubbles. One key observation is that, before the appearance of large bubbles, the liquid films between the bubbles thin out, resulting in a drier foam structure as the shear rate increases for the Foam 1. This idea is further supported by Fig.~\ref{snapshots}(b), which shows the measured liquid content as a function of the imposed shear rate in these experiments. In contrast, the Foam 2, under otherwise similar conditions, shows no signs of foam drying or coalescence, maintaining its structure throughout the experiment. The decline in shear stress for Foam 1 is presumably due to flow-induced bubble coalescence. Note that under all other boundary conditions, the foam structure remained unchanged.


\HM{As shown in Fig.~\ref{FC}(a), introducing roughened conditions generally increases the shear stress at the top wall of the geometry. We hypothesize that this effect arises from reduced wall slip, which in turn enhances shear stress transmission. Liquid foams are known to exhibit wall slip on smooth surfaces~\cite{omirbekov2020experimental,marze2008aqueous,herzhaft1999rheology}. To quantify this, we measured the difference between the imposed wall velocity and the bubble layer velocity. Our results indicate that wall slip remains consistent regardless of smooth boundaries or imposed shear rates (see Fig. S5 in the supplementary materials). For roughened surfaces, direct wall-slip measurements are not feasible due to surface opacity; however, it is expected to be significantly reduced compared to smooth surfaces. Since wall slip lowers the effective shear rate, its reduction leads to an increase in shear stress. The experiments reported in this paper are performed on a single bubble layer sandwitched between two plates, therefore, the shear stress is dominated by the viscous friction between the wall and bubble monolayer. The viscous friction between wall and the bubble layer can be described by Bretherton-type friction law~\cite{bretherton1961motion}. For such conditions,} and a mobile interface, it is predicted that wall shear stress $\tau_w\propto$ Ca$_w^{2/3}$ and for an immobile interface $\tau_w\propto$ Ca$_w^{1/2}$. Here $\tau_w = \tau_v R/\sigma$, and Ca$_w = \mu V_w/\sigma$ is a wall capillary number and $V_w$ is the wall velocity. 
\begin{figure}[hthp]
\centering
\includegraphics[width=0.49\linewidth]{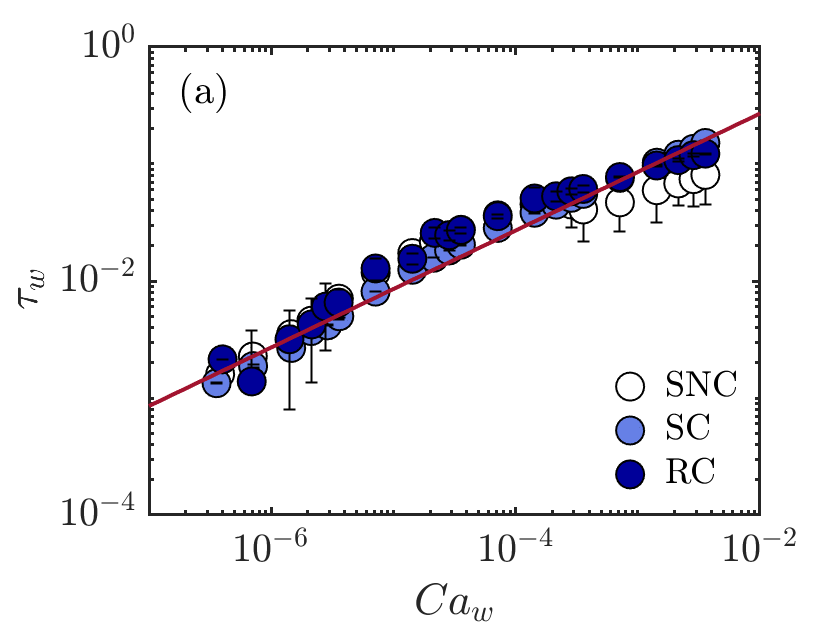}
\includegraphics[width=0.49\linewidth]{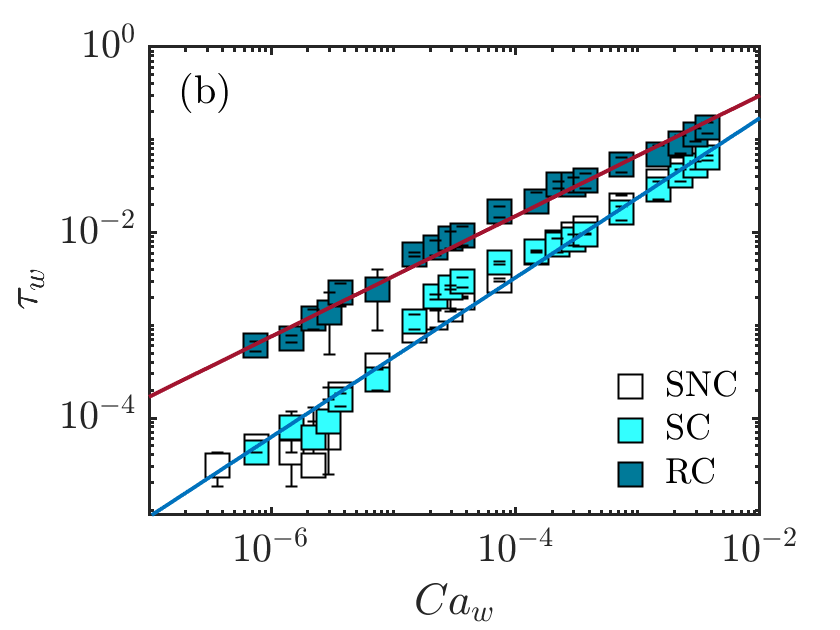}
\caption{\small Dimensionless foam-wall friction stress as function of wall capillary number under various boundary conditions for Foam 1 (a) and Foam 2 (b). The slope of the fitted line in part (a) is 0.5 and in (b) are 0.65 and 0.86.} 
\label{viscous stress}
\end{figure}
Fig.~\ref{viscous stress} presents the wall viscous stress as a function of the wall capillary number for liquid foams based on Foam 1 and Foam 2 under various boundary conditions. For Foam 1 samples, the flow curves of different boundary conditions overlap, following a scaling of $\tau_w \propto$ Ca$_w^{0.5}$. \HM{To our knowledge, a similar scaling has not been previously reported for 2D liquid foams with immobile interfaces. Most existing studies on 2D liquid foams, particularly under bubble-wall friction-dominated conditions, have focused on surfactants that produce mobile interfaces, yielding a characteristic power-law index of $n \approx 2/3$~\cite{katgert2008rate,katgert2010couette}. Thus, our measurements for immobile interfaces in 2D liquid foams are the first of their type. Interestingly, a comparable power-law index of $n \approx 0.5$ has been observed in the flow of 3D liquid foams with immobile interfaces~\cite{denkov2005wall,golemanov2008surfactant}. In contrast, Foam 2 exhibits distinct scaling behavior: for smooth boundaries, $\tau_w \propto$ Ca$_w^{0.86}$, while for rough boundaries, $\tau_w \propto$ Ca$_w^{0.65}$. The rough boundary scaling aligns with prior findings on 2D and 3D foams with mobile interfaces~\cite{katgert2008rate,katgert2010couette}, but the smooth boundary scaling exceeding $2/3$ has not, to our knowledge, been previously reported in experiments. It has been suggested that the viscosity of the surfactant solution may affect the power-law index of the 3D liquid foams~\cite{denkov2009role}. The measured shear viscosities of both soap solutions are similar, 1 mPa.s, ruling out differences in solution viscosity as the cause for the higher power-law exponent. According to the literature, a power-law index greater than $1/2$ is typically observed in liquid foams with a liquid content above 25$\%$ (air volume fraction $<$75$\%$), where bubble-bubble interactions dominate~\cite{denkov2005wall}. Although wall-bubble friction is dominant in our experiments, the liquid fraction is well below 25$\%$, which rules out the latter possibility on the impact of liquid fraction on the power-law index. Therefore, it is still unclear what gives rise to a power-law index higher than 2/3 in Foam 2. On the other hand, the comparison of power-law exponents for Foam 1 and Foam 2, where $ n_{\text{Foam 1}} < n_{\text{Foam 2}}$ under rough boundary conditions indicates that the interface of Foam 1 is less mobile than that of Foam 2. This finding aligns with the linear viscoelastic rheology analysis presented earlier, further supporting the distinction in interfacial properties between the two foams.} \HM{Many household products (e.g., laundry detergent, dishwashing liquid, shaving cream, and toothpaste) are designed to generate foams during use and rely on diverse surfactant chemistries, adding complexity to product formulation. Our findings demonstrate that rheology differentiates foam properties based on surfactant chemistry, underscoring its potential as a powerful tool for optimizing and tailoring product performance.}
\section{Conclusion and Outlook}
In summary, we investigated the flow and structure of 2D liquid foams in both linear and non-linear viscoelastic regimes, using commercially available soap solutions based on Foam 1 and Foam 2 detergents under smooth and rough boundary conditions. The Foam 1 consistently exhibited higher elastic and loss moduli and a lower viscoelastic relaxation frequency compared to that of the Foam 2, suggesting that the liquid-air interface in the Foam 1 is less mobile. However, interfacial modulus measurements revealed that the dilatation modulus of the air-liquid interface was weaker for Foam 1, contradicting the linear viscoelastic data. \HM{Our surface tension measurements confirm that the detergent concentration used in this study is below the CMC, indicating that interfacial rheology measurements are influenced by Marangoni stresses. These stresses introduce additional complexity, further complicating direct comparisons with linear viscoelasticity results.}\par 

Non-linear rheology experiments showed further differences. The viscous stress was higher in the Foam 1, while the Foam 2 exhibited yield stress behavior under smooth boundary conditions, which the Foam 1 did not, despite its higher viscous stress. Additionally, the Foam 1 displayed flow-induced coalescence, leading to a reduction in shear stress at moderately high shear rates. The viscous friction measurements indicated that $\tau_w \propto$ Ca$_w^{0.5}$ for all boundary conditions in the Foam 1, while for the Foam 2, $\tau_w \propto$ Ca$_w^{0.86}$ under smooth and $\tau_w \propto$ Ca$_w^{0.65}$ under rough boundary conditions. \HM{Overall, the linear and non-linear viscoelastic measurements} suggest that the Foam 1 has a less mobile interface than the Foam 2, and roughened boundaries enhance both the elastic and viscous properties in both liquid foams.\par 

\HM{Future studies should further explore the connection between foam rheology and interfacial rheology across a wide range of surfactant concentrations, encompassing both below and beyond the CMC. Additionally, expanding 2D foam flow studies to include a variety of surfactant systems would be valuable in assessing the power-law index for different surfactant chemistries.}\par 

\section{Acknowledgment}
This work is financially supported by Colgate-Palmolive company. We are grateful to Jonghun Lee for his help with the oscillating pendant drop experiments. 

\bibliographystyle{unsrt}
\bibliography{sample}

\begin{thebibliography}{10}

\bibitem{dollet2014rheology}
Benjamin Dollet and Christophe Raufaste.
\newblock Rheology of aqueous foams.
\newblock {\em Comptes Rendus Physique}, 15(8-9):731--747, 2014.

\bibitem{cohen2013flow}
Sylvie Cohen-Addad, Reinhard H{\"o}hler, and Olivier Pitois.
\newblock Flow in foams and flowing foams.
\newblock {\em Annual Review of Fluid Mechanics}, 45(1):241--267, 2013.

\bibitem{denkov2020physicochemical}
Nikolai Denkov, Slavka Tcholakova, and Nadya Politova-Brinkova.
\newblock Physicochemical control of foam properties.
\newblock {\em Current Opinion in Colloid \& Interface Science}, 50:101376, 2020.

\bibitem{hohler2005rheology}
Reinhard H{\"o}hler and Sylvie Cohen-Addad.
\newblock Rheology of liquid foam.
\newblock {\em Journal of Physics: Condensed Matter}, 17(41):R1041, 2005.

\bibitem{conn2014visualizing}
Charles~A Conn, Kun Ma, George~J Hirasaki, and Sibani~Lisa Biswal.
\newblock Visualizing oil displacement with foam in a microfluidic device with permeability contrast.
\newblock {\em Lab on a Chip}, 14(20):3968--3977, 2014.

\bibitem{wei2018foam}
Peng Wei, Wanfen Pu, Lin Sun, Wei Zhou, and Xudong Ji.
\newblock Foam stabilized by alkyl polyglycoside and isoamyl alcohol for enhancing oil recovery in the low-permeable reservoir.
\newblock {\em Journal of Petroleum Science and Engineering}, 171:1269--1278, 2018.

\bibitem{salonen2012dual}
Anniina Salonen, Romain Lhermerout, Emmanuelle Rio, Dominique Langevin, and Arnaud Saint-Jalmes.
\newblock Dual gas and oil dispersions in water: production and stability of foamulsion.
\newblock {\em Soft Matter}, 8(3):699--706, 2012.

\bibitem{somasundaran1972foam}
Ponisseril Somasundaran.
\newblock Foam separation methods.
\newblock {\em Separation and Purification methods}, 1(1):117--198, 1972.

\bibitem{arzhavitina2010foams}
Alexandra Arzhavitina and Hartwig Steckel.
\newblock Foams for pharmaceutical and cosmetic application.
\newblock {\em International journal of pharmaceutics}, 394(1-2):1--17, 2010.

\bibitem{parsa2019foam}
Maryam Parsa, Anna Trybala, Danish~Javed Malik, and Victor Starov.
\newblock Foam in pharmaceutical and medical applications.
\newblock {\em Current Opinion in Colloid \& Interface Science}, 44:153--167, 2019.

\bibitem{zocchi2008foam}
Germaine Zocchi.
\newblock Foam in consumer products.
\newblock {\em Handbook of Detergents. Part A: Properties}, page 419, 2008.

\bibitem{durian1991scaling}
Douglas~J Durian, DA~Weitz, and DJ~Pine.
\newblock Scaling behavior in shaving cream.
\newblock {\em Physical Review A}, 44(12):R7902, 1991.

\bibitem{durian1995foam}
Douglas~J Durian.
\newblock Foam mechanics at the bubble scale.
\newblock {\em Physical review letters}, 75(26):4780, 1995.

\bibitem{cohen2001bubble}
Sylvie Cohen-Addad and Reinhard H{\"o}hler.
\newblock Bubble dynamics relaxation in aqueous foam probed by multispeckle diffusing-wave spectroscopy.
\newblock {\em Physical Review Letters}, 86(20):4700, 2001.

\bibitem{saint2006physical}
Arnaud Saint-Jalmes.
\newblock Physical chemistry in foam drainage and coarsening.
\newblock {\em Soft Matter}, 2(10):836--849, 2006.

\bibitem{weaire1999physics}
Denis~L Weaire and Stefan Hutzler.
\newblock {\em The physics of foams}.
\newblock Oxford University Press, 1999.

\bibitem{denkov2012foam}
Nikolai~D Denkov, Slavka~S Tcholakova, Reinhard H{\"o}hler, and Sylvie Cohen-Addad.
\newblock Foam rheology.
\newblock {\em Foam engineering: Fundamentals and applications}, pages 91--120, 2012.

\bibitem{reinelt1990shearing}
Douglas~A Reinelt and Andrew~M Kraynik.
\newblock On the shearing flow of foams and concentrated emulsions.
\newblock {\em Journal of Fluid Mechanics}, 215:431--455, 1990.

\bibitem{ovarlez2010investigation}
Guillaume Ovarlez, Kapil Krishan, and Sylvie Cohen-Addad.
\newblock Investigation of shear banding in three-dimensional foams.
\newblock {\em Europhysics Letters}, 91(6):68005, 2010.

\bibitem{langevin2014rheology}
Dominique Langevin.
\newblock Rheology of adsorbed surfactant monolayers at fluid surfaces.
\newblock {\em Annual review of fluid mechanics}, 46(1):47--65, 2014.

\bibitem{buzza1995linear}
David~Martin Buzza, C-YD Lu, and Michael Cates.
\newblock Linear shear rheology of incompressible foams.
\newblock {\em Journal de Physique II}, 5(1):37--52, 1995.

\bibitem{denkov2009role}
Nikolai~D Denkov, Slavka Tcholakova, Konstantin Golemanov, KP~Ananthpadmanabhan, and Alex Lips.
\newblock The role of surfactant type and bubble surface mobility in foam rheology.
\newblock {\em Soft Matter}, 5(18):3389--3408, 2009.

\bibitem{marze2008aqueous}
Sebastien Marze, Dominique Langevin, and Arnaud Saint-Jalmes.
\newblock Aqueous foam slip and shear regimes determined by rheometry and multiple light scattering.
\newblock {\em Journal of rheology}, 52(5):1091--1111, 2008.

\bibitem{golemanov2008breakup}
Konstantin Golemanov, S~Tcholakova, Nikolai~D Denkov, KP~Ananthapadmanabhan, and Alex Lips.
\newblock Breakup of bubbles and drops in steadily sheared foams and concentrated emulsions.
\newblock {\em Physical Review E—Statistical, Nonlinear, and Soft Matter Physics}, 78(5):051405, 2008.

\bibitem{denkov2005wall}
Nikolai~D Denkov, Vivek Subramanian, Daniel Gurovich, and Alex Lips.
\newblock Wall slip and viscous dissipation in sheared foams: Effect of surface mobility.
\newblock {\em Colloids and Surfaces A: Physicochemical and Engineering Aspects}, 263(1-3):129--145, 2005.

\bibitem{golemanov2008surfactant}
Konstantin Golemanov, Nikolai~D Denkov, Slavka~S Tcholakova, M~Vethamuthu, and Alex Lips.
\newblock Surfactant mixtures for control of bubble surface mobility in foam studies.
\newblock {\em Langmuir}, 24(18):9956--9961, 2008.

\bibitem{bretherton1961motion}
Francis~Patton Bretherton.
\newblock The motion of long bubbles in tubes.
\newblock {\em Journal of Fluid Mechanics}, 10(2):166--188, 1961.

\bibitem{schwartz1986motion}
Leonard Schwartz, HM~Princen, and AD~Kiss.
\newblock On the motion of bubbles in capillary tubes.
\newblock {\em Journal of Fluid Mechanics}, 172:259--275, 1986.

\bibitem{debregeas2001deformation}
Georges Debregeas, Herve Tabuteau, and J-M Di~Meglio.
\newblock Deformation and flow of a two-dimensional foam under continuous shear.
\newblock {\em Physical Review Letters}, 87(17):178305, 2001.

\bibitem{lauridsen2004velocity}
John Lauridsen, Gregory Chanan, and Michael Dennin.
\newblock Velocity profiles in slowly sheared bubble rafts.
\newblock {\em Physical Review Letters}, 93(1):018303, 2004.

\bibitem{katgert2008rate}
Gijs Katgert, Matthias~E M{\"o}bius, and Martin van Hecke.
\newblock Rate dependence and role of disorder in linearly sheared two-dimensional foams.
\newblock {\em Physical review letters}, 101(5):058301, 2008.

\bibitem{katgert2010couette}
Gijs Katgert, Brian~P Tighe, Matthias~E M{\"o}bius, and Martin van Hecke.
\newblock Couette flow of two-dimensional foams.
\newblock {\em Europhysics Letters}, 90(5):54002, 2010.

\bibitem{dennin2004statistics}
Michael Dennin.
\newblock Statistics of bubble rearrangements in a slowly sheared two-dimensional foam.
\newblock {\em Physical Review E—Statistical, Nonlinear, and Soft Matter Physics}, 70(4):041406, 2004.

\bibitem{wang2006impact}
Yuhong Wang, Kapilanjan Krishan, and Michael Dennin.
\newblock Impact of boundaries on velocity profiles in bubble rafts.
\newblock {\em Physical Review E—Statistical, Nonlinear, and Soft Matter Physics}, 73(3):031401, 2006.

\bibitem{mohammadigoushki2012anomalous}
Hadi Mohammadigoushki, Giovanni Ghigliotti, and James~J Feng.
\newblock Anomalous coalescence in sheared two-dimensional foam.
\newblock {\em Physical Review E}, 85(6):066301, 2012.

\bibitem{mohammadigoushki2013size}
Hadi Mohammadigoushki and James~J Feng.
\newblock Size segregation in sheared two-dimensional polydisperse foam.
\newblock {\em Langmuir}, 29(5):1370--1378, 2013.

\bibitem{mader2012quantitative}
Kevin Mader, Rajmund Mokso, Christophe Raufaste, Benjamin Dollet, St{\'e}phane Santucci, J{\'e}r{\^o}me Lambert, and Marco Stampanoni.
\newblock Quantitative 3d characterization of cellular materials: Segmentation and morphology of foam.
\newblock {\em Colloids and Surfaces A: Physicochemical and Engineering Aspects}, 415:230--238, 2012.

\bibitem{mohammadigoushki2015temporal}
Hadi Mohammadigoushki and James~J Feng.
\newblock Temporal evolution of microstructure and rheology of sheared two-dimensional foams.
\newblock {\em Journal of Non-Newtonian Fluid Mechanics}, 223:1--8, 2015.

\bibitem{janiaud2006two}
Eric Janiaud, Denis Weaire, and Stefan Hutzler.
\newblock Two-dimensional foam rheology with viscous drag.
\newblock {\em Physical review letters}, 97(3):038302, 2006.

\bibitem{langlois2008rheological}
Vincent~J Langlois, Stefan Hutzler, and Denis Weaire.
\newblock Rheological properties of the soft-disk model of two-dimensional foams.
\newblock {\em Physical Review E—Statistical, Nonlinear, and Soft Matter Physics}, 78(2):021401, 2008.

\bibitem{krishan2010fast}
Kapilanjan Krishan, Ahmed Helal, Reinhard H{\"o}hler, and Sylvie Cohen-Addad.
\newblock Fast relaxations in foam.
\newblock {\em Physical Review E—Statistical, Nonlinear, and Soft Matter Physics}, 82(1):011405, 2010.

\bibitem{costa2013coupling}
S{\'e}verine Costa, Reinhard H{\"o}hler, and Sylvie Cohen-Addad.
\newblock The coupling between foam viscoelasticity and interfacial rheology.
\newblock {\em Soft Matter}, 9(4):1100--1112, 2013.

\bibitem{mohammadigoushki2014bubble}
Hadi Mohammadigoushki, Pengtao Yue, and James~J Feng.
\newblock Bubble migration in two-dimensional foam sheared in a wide-gap couette device: Effects of non-newtonian rheology.
\newblock {\em Journal of Rheology}, 58(6):1809--1827, 2014.

\bibitem{liu1996anomalous}
Andrea~J Liu, Sriram Ramaswamy, TG~Mason, Hu~Gang, and DA~Weitz.
\newblock Anomalous viscous loss in emulsions.
\newblock {\em Physical review letters}, 76(16):3017, 1996.

\bibitem{langevin2014marangoni}
Dominique Langevin and Francisco Monroy.
\newblock Marangoni stresses and surface compression rheology of surfactant solutions. achievements and problems.
\newblock {\em Advances in colloid and interface science}, 206:141--149, 2014.

\bibitem{marquez2025measurement}
Ronald Marquez and Jean-Louis Salager.
\newblock Measurement techniques for interfacial rheology of surfactant, asphaltene, and protein-stabilized interfaces in emulsions and foams.
\newblock {\em Colloids and Interfaces}, 9(1):14, 2025.

\bibitem{omirbekov2020experimental}
Sagyn Omirbekov, Hossein Davarzani, St{\'e}fan Colombano, and Azita Ahmadi-Senichault.
\newblock Experimental and numerical upscaling of foam flow in highly permeable porous media.
\newblock {\em Advances in Water Resources}, 146:103761, 2020.

\bibitem{herzhaft1999rheology}
Benjamin Herzhaft.
\newblock Rheology of aqueous foams: a literature review of some experimental works.
\newblock {\em Oil \& gas science and technology}, 54(5):587--596, 1999.

\end{thebibliography}

\end{document}